\documentclass[twocolumn,preprintnumbers,superscriptaddress,nofootinbib,showpacs,showkeys]{revtex4}
\usepackage{amsmath}
\usepackage{amssymb}
\usepackage{graphicx}
\usepackage{hyperref}
\usepackage{xspace}
\usepackage{bm}
\usepackage{slashed}
\usepackage{dsfont}
\usepackage{bbm}

\usepackage{booktabs}
\AtBeginDocument{
\heavyrulewidth=.08em
\lightrulewidth=.05em
\cmidrulewidth=.03em
\belowrulesep=.65ex
\belowbottomsep=0pt
\aboverulesep=.4ex
\abovetopsep=0pt
\cmidrulesep=\doublerulesep
\cmidrulekern=.5em
\defaultaddspace=.5em
}

\newcommand{\be}{\begin{equation}}
\newcommand{\ee}{\end{equation}}
\newcommand{\bea}{\begin{eqnarray}}
\newcommand{\eea}{\end{eqnarray}}

\usepackage{xcolor}
\definecolor{light-gray}{gray}{0.8}

\begin{document}

\title{The New Pentaquarks in the Diquark Model}


\newcommand{\CERNaff}{CERN, PH-TH, CH-1211 Geneva 23, Switzerland}
\newcommand{\sapienza}{Dipartimento di Fisica and INFN, `Sapienza' Universit\`a di Roma\\
P.le Aldo Moro 5, I-00185 Roma, Italy}
\newcommand{\columbia}{Department of Physics, 538W 120th Street,
Columbia University, New York, NY, 10027, USA}
\newcommand{\pavia}{INFN Pavia, Via A. Bassi 6, I-27100 Pavia, Italy}
\newcommand{\romadue}{Dipartimento di Fisica and INFN, Universit\`a di Roma `Tor Vergata'\\
 Via della Ricerca Scientifica 1, I-00133 Roma, Italy}
\newcommand{\alice}{ALICE\xspace}

\author{L. Maiani }
\affiliation{\sapienza}
\author{A.D.~Polosa}
\affiliation{\sapienza}
\author{V.~Riquer}
\affiliation{\sapienza}

\begin{abstract}
Pentaquark baryons are a natural expectation of an extended picture of hadrons where quarks and diquarks are the fundamental units. The parity/mass pattern observed, when compared to that of exotic mesons, appears as the footprint of  a compact five-quark structure. What has been learned from the $X,Y,Z$ phenomenology informs about the newly found pentaquark structure and suggests further experimental tests 
and directions to be explored.
 \end{abstract}


\maketitle

\section*{Introduction}
The LHCb collaboration has reported observation of two new resonances 
in the $\Lambda_b$ decay~\cite{penta},
\be
\Lambda_b (bud)\to \mathds{P}^+ K^-
\label{uno}
\ee
 each decaying according to
 \be
\mathds{P}^+ \to J/\Psi + p
\label{due}
\ee
Thus the new particles carry a unit of baryonic number and feature the valence quark composition 
 \be
\mathds{P}^+ = \bar c c u u d
\label{valence}
\ee
 whence the name pentaquarks.
 
The  best fit quantum numbers and masses are~\footnote{We refer to the original article~\cite{penta} for experimental errors and more details.}
 \bea
 &&J^P=~3/2^-,M\simeq 4380~{\rm GeV},~{\rm fract.}\simeq 8.4~\%\nonumber \ \\
 &&J^P=~5/2^+,M\simeq 4450~{\rm GeV},~{\rm fract.}\simeq 4.1~\%  
 \label{5/2}
 \eea
 
In this note, we comment on the two pentaquarks as the logical extension of the picture already proposed in~\cite{noi1}, and for the beauty sector in~\cite{ali}, for the exotic mesons, $X,Y,Z$, whereby the latter particles are described using diquarks as colored subunits, bound by QCD color forces. See also the discussions in~\cite{wbrod}.

Light scalar mesons as four quark states have been considered in~\cite{jaffe1} and further studied in~\cite{{Maiani:2004uc},hooft}. Heavy-light diquarks as building blocks of  hidden charm or beauty exotic mesons have been introduced in~\cite{noi1,ali}. 
Pentaquarks from light diquarks are described in~\cite{jw,Rossi:2004yr} see also~\cite{jaffe}.
Hidden charm pentaquarks were anticipated in~\cite{lm}.

In the particular case of the newly discovered pentaquarks, we are led to identify the basic (color $\bar{\bm 3}$) units as: the charm antiquark $\bar c$, one heavy-light diquark, $[cq]$, and one light-light diquark, $[q^\prime q^{\prime\prime}]$ ($q, q^\prime, q^{\prime\prime}$ denote light quarks, which we restric at first to be the $u,d$ quarks, extending later to the flavor SU(3) triplet, $u,d,s$).

Needless to say, the picture of colored sub-units opens the door to a rich spectroscopy of  states, including orbital excitations in addition to $S$-wave states, not dissimilar from the baryon spectrum, with the $\bm{56}$ positive parity baryons followed by the $\bm{70},~L=1$  multiplet of negative parity baryons. 

A precise description of pentaquark spectroscopy has to wait for more particles to be identified. However, we shall see that even the two states just observed carry enough information to corroborate the diquark role in the new baryons and mesons and lead to identify some crucial experimental signature that could make decisive progress in this matter.

\section*{Pentaquark Parity}

Light, $S$-wave mesons have negative parity, being made by a quark-antiquark pair whose components have opposite parity. Negative parity  are followed by positive parity states ($A_{1,2}$, $\chi_J$ states, etc.) due to the excitation of one unit of orbital angular momentum. The negative parity of the lighter state in~(\ref{5/2}) reflects just the presence of one valence antiquark in (\ref{valence}) and the positive parity of the  next state is naturally interpreted as the opening of the orbital, $L=1$, excitation. Parity ordering in the baryons, that we have just recalled, and in $X,Y,Z$ mesons, is just the opposite, the $X(3872)$ with $J^{PC}=1^{++}$, being lighter  than $Y(4260)$, with $J^{PC}=1^{--} $. This feature, of course, reflects the fact that there are  no valence antiquarks in the familiar baryons and two quark-antiquark pairs in the lowest lying $X, Z$ mesons, as required by the tetraquark picture.

\section*{The mass difference}

At first sight, the near $70$~MeV difference between the masses in~(\ref{5/2})  does not go well with the energy associated to orbital excitation. One orbital excitation in mesons and baryons carries an energy difference which is typically of order $300$ MeV, as exemplified by the  mass difference in $\Lambda(1405)-\Lambda(1116)\sim 290$~MeV. Mass formulae for the orbital excitation in  $X, Y, Z$ mesons are discussed in~\cite{type2} and the associated energy difference is estimated to be $\Delta M ( L=0\to 1)\sim 280$~MeV. 

However, the mass difference between light-light diquarks with spin $s=1,0$~\cite{DeRujula}, estimated from charm and beauty baryon spectra, is of order $200$~MeV, {\it e.g.} $\Sigma_c(2455)-\Lambda_c(2286)\simeq 170$~MeV, $\Sigma_b(5811)-\Lambda_b(5620)\simeq 190$~MeV. 

If we assume the compositions
\bea
&&\mathds{P}( 3/2^-) = \left\{\bar c\, [cq]_{s=1} [q^\prime q^{\prime \prime}]_{s=1}, L=0\right\}\nonumber 
\\
&&\mathds{P}( 5/2^+) =\left\{\bar c\, [cq]_{s=1} [q^\prime q^{\prime \prime}]_{s=0}, L=1\right\} 
\label{compos} 
\eea
the orbital gap is reduced to about $100$~MeV, which brings it back to the range of spin-spin  and spin-orbit corrections indicated by (\ref{5/2}). 

\begin{figure*}[t]
 \begin{center}
   \includegraphics[width=\columnwidth]{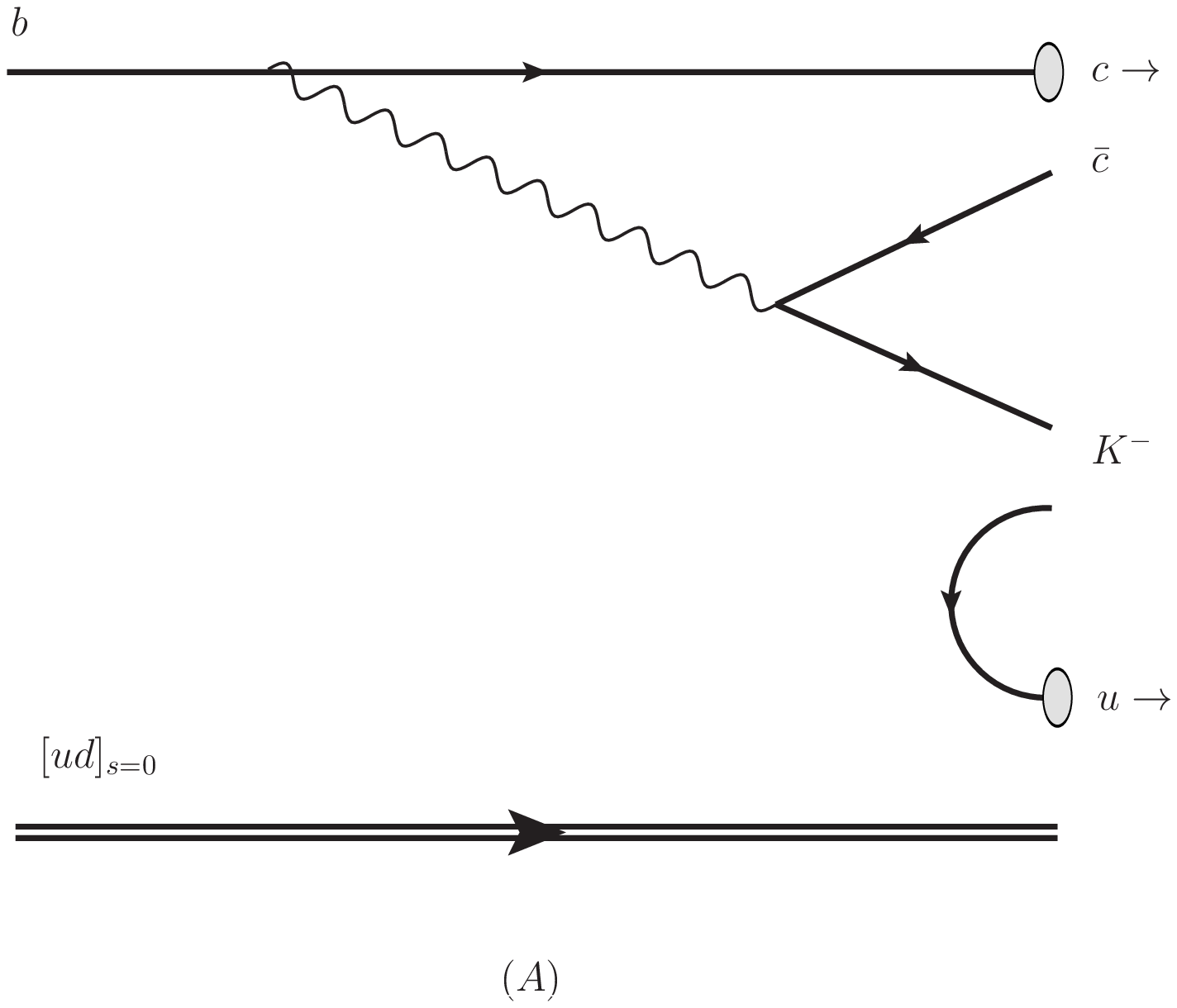} \hspace{.2cm} \includegraphics[width=\columnwidth]{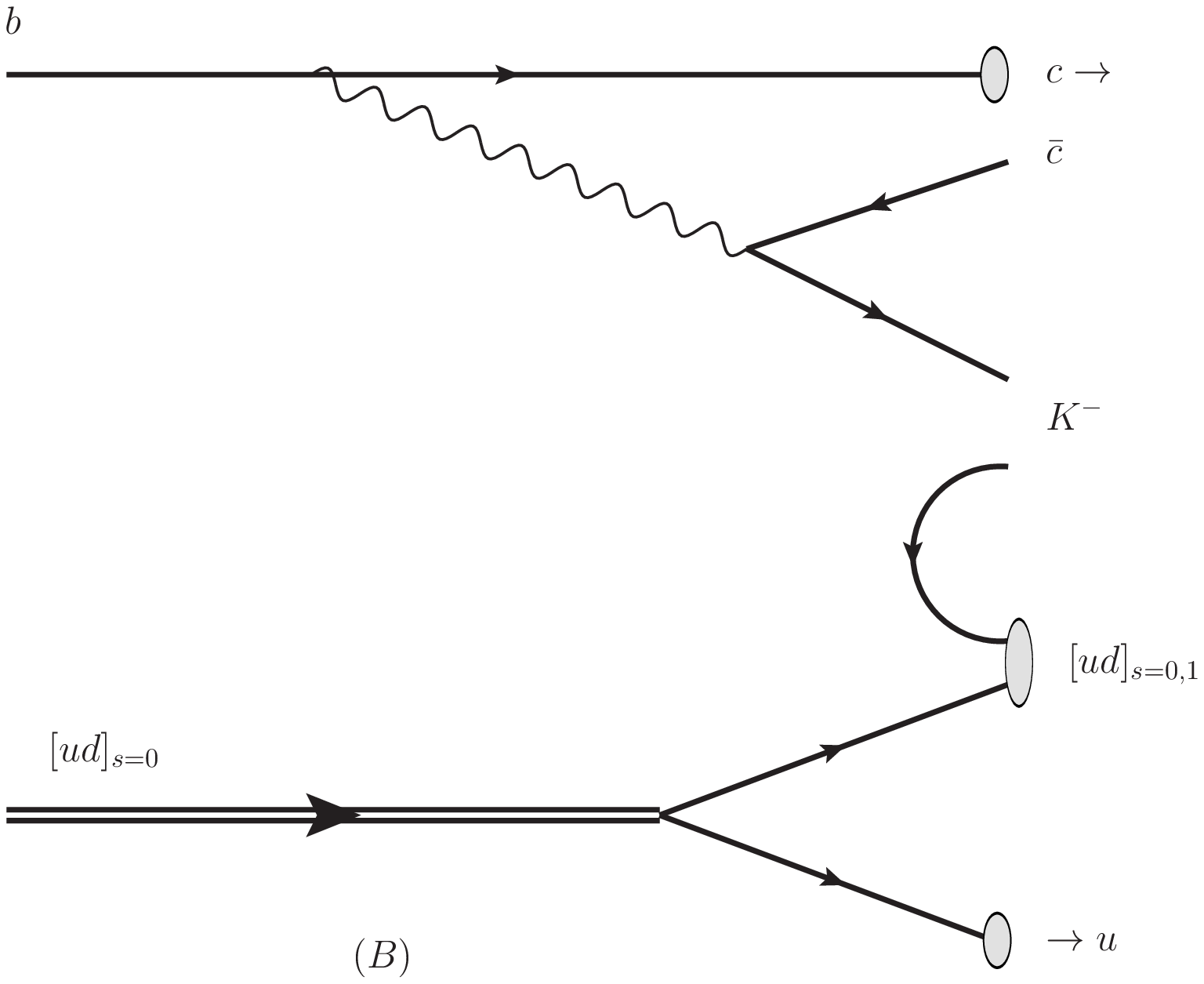}
 \end{center}
\caption{\small $(A)$: The $[ud]$, spin zero diquark in the  $\Lambda_b$  is transmitted to the $\mathds{P}_u$ type pentaquark. $(B)$ The $u$ quark from the vacuum particpates in the formation of the light-light diquark: spin zero and one are both permitted. Mechanism (B) may also produce a $[uu]_{s=1}$ diquark, giving rise to the $\mathds{P}_d$ type pentaquark, Eq.~(\ref{pentad}). \label{funo}}
\end{figure*}

Spin one light-light diquarks (Jaffe's bad diquarks~\cite{jaffe}), while conspicuously absent in light meson spectroscopy, are well established in baryons as indicated by  the $\Sigma-\Lambda$ mass diference~\cite{DeRujula} and confirmed by $\Sigma_{c,b}-\Lambda_{c,b}$ mass differences~\cite{noi1}. 

Concernig the production of a spin 1, light-light $[ud]$ diquark in  $\Lambda_b$ decay, we note that there are in fact two possible mechanisms leading to the pentaquark production, see Fig.~\ref{funo}. In the first one (diagram $A$ in Fig.~\ref{funo}) the $b$-quark spin is shared between the Kaon and the $\bar c$ and $[cu]$ components. Barring angular momentum transfer due to gluon exchanges between the light diquark and light quarks from the vacuum, the final $[ud]$ diquarks has to have spin zero. In the second mechanism, however (diagram $B$), the $[ud]$ diquark is formed from the original $d$ quark and the $u$ quark from the vacuum. Angular momentum is shared among all final components and the  $[ud]$ diquark may well have spin one. The two possibilities are considered in the following discussion.

Concerning heavy quark spin conservation, one can also  show that both $\mathds{P}( 3/2^-)$ and $\mathds{P}( 5/2^+)$, as described in~(\ref{compos}), have components with $s_{c\bar c}=1$. This can be seen by direct inspection or with an SU(2) tensor analysis. HQS conservation in pentaquark decay allows the production of $J/\Psi $ in the final state, as observed.

\section*{Flavor SU(3) structure of pentaquarks}

Pentaquarks realizing the valence quark structure (\ref{valence}) are of two types
\bea
\label{pentau} 
&&\mathds{P}_u=\epsilon^{\alpha\beta\gamma}\;\bar c_\alpha\, [cu]_{\beta,s=0,1}\, [ud]_{\gamma,s=0,1}\\
&&\mathds{P}_d=\epsilon^{\alpha\beta\gamma}\;\bar c_\alpha\, [cd]_{\beta,s=0,1}\, [uu]_{\gamma,s=1}
\label{pentad}
\eea
where greek indices are for color, diquarks are in the color  antisymmetric, $\bar{\bm 3}$, configuration and overall antisymmetry requires flavor symmetric light-light diquark with $s= 1$.

Extending to flavor SU(3), we have two distinct series of pentaquarks according the light-light diquark symmetry
\bea
\mathds{P}_A &=&\epsilon^{\alpha\beta\gamma}\;\{\bar c_\alpha\, [cq]_{\beta,s=0,1}\, [q^\prime q^{\prime\prime}]_{\gamma,s=0},  L\}  = \nonumber \\
 &=& {\bm 3}\otimes  \bar{\bm 3}= {\bm 1}\oplus {\bm 8}
\label{nonet}\\
\mathds{P}_S &=&\epsilon^{\alpha\beta\gamma}\;\{\bar c_\alpha\, [cq]_{\beta,s=0,1}\, [q^\prime q^{\prime\prime}]_{\gamma,s=1},L \} =\nonumber \\
&=& {\bm 3}\otimes {\bm 6} = {\bm 8}\oplus {\bm{10}}
\label{octodec}
\eea

For $S$-waves, the first and the second series give the angular momenta
\bea
&&\mathds{P}_A(L=0):~J=1/2~(2), 3/2~(1)\label{antis}\\
&&\mathds{P}_S(L=0):~J=1/2~(3), 3/2~(3), 5/2~(1)\label{sim}
\eea
(in parenthesis the multiplicity of each spin value). In consideration of~(\ref{compos}), we propose to assign the $3/2^-$ and the $5/2^+$ states to the symmetric and antisymmetric serieses, respectively.

To study the flavor properties of pentaquark production and decay, we recall that 
\be
\Lambda_b(bud) \sim {\bar {\bm 3}}
\ee
with respect to flavor SU(3) and is isosinglet $I=0$.
The weak non-leptonic Hamiltonian for $b$ decay is\footnote{We denote strangeness by $S$, not to be confused with the diquark spin $s$.}
\be
H_w^{({\bm 3})}(\Delta I=0,~\Delta S=-1)
\ee
Therefore, denoting by $M$ a nonet light meson, the weak transition amplitude 
\be
\langle \mathds{P},M|H_w|\Lambda_b\rangle
\label{decay1}
\ee
requires $\mathds{P}+M$ to be in the ${\bm 8}\oplus {\bm 1}$  representation. Recalling the well known SU(3) formulae
\bea
&&{\bm 8}\otimes {\bm 8} ={\bm 1}\oplus {\bm 8}\oplus {\bm 8}\oplus {\bm{10}}\oplus {\overline{{\bm{10}}}}\oplus {\bm{27}} \nonumber \\
&& {\bm 8}\otimes {\bm{10}} ={\bm 8}\oplus {\bm{10}}\oplus {\bm{27}}\oplus {\bm{35}}
\eea
we see that the decay (\ref{decay1}) can be realized with $\mathds{P}$ in either octet or decuplet. The first case is exemplified in Eqs. (\ref{uno}) and (\ref{due}). However, decays such as
\bea
&&\Lambda_b \to \pi \, \mathds{P}_{\bm{10}}^{S=-1}\to \pi\, (J/\Psi \, \Sigma(1385))\nonumber \\
&&\Lambda_b\to K^+ \, \mathds{P}_{\bm{10}}^{S=-2} \to K^+\, (J/\Psi \, \Xi^-(1530))
\eea
might also occur when the $[ud]$ diquark shell in the initial state gets broken in the decay (see $B$ in Fig.~\ref{funo}). 
 
The $\Xi^0_b(bus), \Xi^-(bds)$ and $\Omega_b(bss)$ particles undergo visible weak decays. 
 Example of weak decays from bottom strange baryons involving pentaquarks in the $\bm{10}$ and respecting $\Delta I=0$ and $\Delta S=-1$ are
 \be
\Xi_b(5794)\to  K\,(J/\Psi\, \Sigma(1385))
\ee
in various charge combinations, which would correspond to the formation of the pentaquarks
\be
\mathds{P}_{\bm{10}}(\bar c\, [cq]_{s=0,1}[q^\prime s]_{s=0,1})
\ee
with $q,q^\prime=u,d$.
The $[ss]$ pair in $\Omega_b$ is in pure $\bm 6$ SU(3) representation (with spin one) and we might expect its decay to produce decuplet pentaquarks in association with kaons, with spectacular experimental signatures.
Examples of pentaquark production in $\Omega_b$ decays are
\bea
&&\Omega^-_b(6049)\to \phi\,(J/\Psi\, \Omega^-(1672))\\
&&\Omega^-_b(6049)\to K\, (J/\Psi\, \Xi(1387))
\eea
which would correspond  respectively to the formation of the following pentaquarks
\bea
&&\mathds{P}_{\bm{10}}^-(\bar c\, [cs]_{s=0,1}[ss]_{s=1})\\
&& \mathds{P}_{\bm{10}}(\bar c\, [cq]_{s=0,1}[ss]_{s=1})
\eea
$q=u,d$. These transitions are obtained assuming that the initial $[ss]$ diquark in $\Omega^-_b$ is left unbroken by the decay process.
More transitions can be found relaxing this condition.

 \section*{Conclusions}
 
 The new pentaquarks, with the parity/mass pattern observed by the LHCb collaboration,  are an evident confirmation that diquarks work as an organizing principle for a new class of hadrons we are observing since the discovery of $X(3872)$, back in 2003. In this note we have highlighted the essential features predicted by the antiquark-diquark-diquark scheme of the pentaquark, which matches the experimental evidence so far obtained. 
 
More such exotic baryons are expected and needed to make reliable hypotheses on the way the interactions in the system are shaping the spectra. Crossing the information on pentaquarks and tetraquarks will likely  be the way towards a definitive assessment of exotic hadron spectroscopy. 

We wish to thank G.C. Rossi for useful discussions on baryonia.


\end{document}